\documentclass[twocolumn,showpacs,preprintnumbers,aps]{revtex4}

\usepackage{graphicx}
\usepackage{dcolumn}
\usepackage{bm}

\def\infintd3r{ \int_{-\infty}^\infty d^3r\,}
\def\intd3r{ \int d^3r\,}

\def\laplace1d{\frac{d^2}{dx^2}}
\def\plaplace1d{\frac{d^2}{d{x'}^2}}

\def\padr2{\frac{\partial^2}{\partial r^2}}

\def\bmu{{\vec \mu}}
\def\SMA{^{\rm SMA}}

\def\bea{\begin{eqnarray}}
\def\eea{\end{eqnarray}}
\def\ben{\begin{equation}}
\def\een{\end{equation}}
\def\benu{\begin{enumerate}}
\def\enu{\end{enumerate}}

\def\sss{\scriptscriptstyle\rm}

\def\1var{(\bx_1...\bx\N)}

\def\br{{\bf r}}

\def\bx{{x}}

\def\x{_{\sss X}}
\def\c{_{\sss C}}

\def\xc{_{\sss XC}}
\def\Hx{_{\sss HX}}
\def\Hxc{_{\sss HXC}}

\def\N{_{\sss N}}

\def\ALDA{^{\rm ALDA}}

\def\sph_int{ {\int d^3 r}}

\def\PRL{Phys. Rev. Letts.\ }

\input rotate

\begin{document}


\title{Excitations in time-dependent density-functional theory}
\author{H. Appel}
\altaffiliation[Also at ]{Department of Physics, Rutgers University, Frelinghuysen Road, Piscataway, NJ 08854}
\author{E.K.U. Gross}
\affiliation{Institut f\"ur Theoretische Physik, Freie Universit\"at Berlin, Arnimallee 14, D-14195 Berlin}
\author{K. Burke}
\affiliation{Department of Chemistry and Chemical Biology, Rutgers University, 610 Taylor Rd, Piscataway, NJ 08854}

\date{\today}

\begin{abstract}
An approximate solution to the time-dependent density functional
theory (TDDFT) response equations for finite systems is
developed, yielding corrections to the single-pole
approximation.  These explain why allowed Kohn-Sham transition frequencies
and oscillator strengths are usually good approximations to the
true values, and why sometimes they are not.  The approximation
yields simple expressions for G\"orling-Levy perturbation
theory results, and a method for estimating expectation
values of the unknown exchange-correlation kernel.
\end{abstract}

\maketitle

Traditional density-functional theory (DFT)
is a popular and efficient method for the calculation of ground-state
properties 
of interacting many electron systems.  While it has long
been the method of choice for solid-state calculations \cite{JG89},
recent improvements in approximate functionals have also
made it popular in quantum chemistry, because of the ability
to handle large systems \cite{Kb99}.  
Within exact ground-state theory, the Kohn-Sham eigenvalues
and orbitals are mathematical constructs,
designed to reproduce the ground-state density.
The one exception 
is that the highest occupied eigenvalue is known 
to equal minus the ionization potential of the system.
For many years, the unoccupied Kohn-Sham levels
of solids have been used to interpret the excitations
in solids, despite the well-known underestimate of the
band gap of insulators in LDA.

There exist several extensions of the basic formalism
that allow extraction of excited-state properties.
One popular approach is via {\em time-dependent} DFT (TDDFT),
in which the interacting system in a time-dependent external
field is mapped exactly to a non-interacting time-dependent Kohn-Sham (KS)
system with the same time-dependent density \cite{RG84}.  
If a weak time-dependent electrical field is considered,
this leads to a Dyson-like response equation for the
exact susceptibility of the interacting electronic system.
Poles of this susceptibility occur at the true transition
frequencies \cite{PGG96}, and the strengths of these poles are related
to oscillator strengths.
Solution of these equations has been implemented
in several quantum chemical packages, and excitation spectra
of many molecules have been calculated and reported
in hundreds of papers (See references in Ref. \cite{MBAG01}).
Once an accurate ground-state Kohn-Sham potential is used,
transition frequencies are typically within about 0.2 eV of
experiment.
Oscillator
strengths are usually good in these calculations (within a factor of 2),
but not always.  There are considerable subtleties in applying TDDFT
to extract the
optical response of solids \cite{01BKBL}.

The beauty of using TDDFT to extract optical spectra
is that it combines moderate accuracy with inexpensive
calculation, just as in the ground-state case.
But even beginning from an {\em exact} ground-state KS potential, the
spectrum still depends on the unknown exchange-correlation (XC)
kernel, i.e., the functional derivative of the
time-dependent XC potential.  This is often approximated
by the crude but reliable adiabatic local-density or
generalized-gradient approximations.
These TDDFT calculations compare very favorably with
configuration-interaction singles (CIS) calculations, the only
alternative that is comparable in computational cost \cite{HH99}.  Higher
level calculations, such as more complete CI,
Bethe-Salpeter, or quantum Monte Carlo \cite{GRML01}, can be made more accurate,
but cost more, limiting their use to smaller systems.
For example, the TDDFT approach has recently been applied to electron-transfer 
problems in biological systems \cite{01HFHH}.

Although TDDFT methodology has been implemented and is being
used widely, understanding of its accuracy and reliability has been slow,
as well as its relation to other methods.  The relation to first-order
G\"orling-Levy perturbation theory has been shown \cite{GS99,BPG00},
as well as the connection with the GW approximation \cite{GS99}.
The extreme case of stretched H$_2$ has recently been studied
by several authors \cite{00CGGG,GGBb00,AGR02}, although this
also represents difficulties for the ground-state theory \cite{PSB95}.
By using a matrix formulation of Casida \cite{C96}, the present
paper shows how, when the excitations of a system are discrete,
an approximate solution, that can be made arbitrarily accurate,
can be used to understand and explain many trends in
the results of TDDFT calculations.

To demonstrate our results, we apply them to the prototype
systems of the He and Be atoms.  We chose these because their
exact ground-state KS potentials are known \cite{UG94}, and because
their lowest allowed transitions exhibit two classes of systems we
are interested in.  In the He atom the 1s to 2p singlet transition
is at 21.22 eV, while the KS transition, i.e., the energy
difference between the 1s and 2p ground-state KS orbital energy levels, is 
21.15 eV, less than 0.1 eV smaller.
On the other hand, in the Be atom, the
2s to 2p level singlet transition is at 5.3 eV, but in the KS case it
is at 3.6 eV.   We explain below the fundamental difference between
these two systems, and why the KS eigenvalues are a good approximation
in the first case, but a poor approximation in the second.
Furthermore, the oscillator strength for the 1s to 2p transition in
the He atom is 0.27, but 0.32 in the exact KS case.
Thus the oscillator strength of the He atom KS system is close
to the true one, but noticeably less close than the transition
frequency.  Finally, the 2s to 2p oscillator strength in the
Be atom is 1.37, but 2.54 in its Kohn-Sham alter ego, far
closer than one would expect, given the error in transition
frequencies.  We explain these facts.

We denote the exact KS eigenvalues as $\epsilon_i$ and
orbitals as $\phi_i(\br)$.
Casida \cite{C96} has written the TDDFT response equations
as an eigenvalue equation for the square of the transition frequencies:
\ben
\sum_{q'} {\tilde \Omega}_{qq'} (\omega) v_{q'} = \Omega v_q,
\label{mat}
\een
where $q$ is a double index, representing a transition from
occupied KS orbital $i$ to unoccupied KS orbital $a$,
$\omega_q=\epsilon_a-\epsilon_i$, $\Omega=\omega^2$, 
and $\Phi_q(\br) = \phi_i^*(\br)\phi_a(\br)$.
The matrix is
\ben
{\tilde \Omega}_{qq'}
= \delta_{qq'} \Omega_q + 2 {\sqrt{\omega_q\omega_q'}} 
\langle q | f\Hxc (\omega) | q' \rangle,
\label{Odef}
\een
where
\ben
\langle q | f\Hxc(\omega) | q' \rangle
= \int d^3r\int d^3r' \Phi^*_q(\br)
f\Hxc(\br,\br',\omega)\Phi_{q'}(\br').
\label{Mdef}
\een
In this equation, $f\Hxc$ is the Hartree-exchange-correlation
kernel, $1/|\br-\br'|+f\xc(\br,\br',\omega)$, where
$f\xc$ is the unknown XC kernel.

It has been noticed that
the KS transition frequencies are often `good' approximations
to the true frequencies \cite{ARU98}.
If the transition frequencies and oscillator strengths are
expanded in powers of the coupling constant $\lambda$, as in GL 
perturbation theory \cite{GL92}, the zero-order values are the 
Kohn-Sham values \cite{G96}.
In Eq. (\ref{Mdef}), we see that,
if $f\Hxc$ is small, i.e., if the system is sufficiently weakly correlated,
this is correct.
We show below that this is {\em not} the reason why the KS values
are good approximations in many systems, such as the He atom, and
is especially untrue when the ground-state has a near-degeneracy,
such as for the Be atom.

Our approximate solution relies on the fact that
$\langle q | f\Hxc | q' \rangle$ decays rapidly
with distance from the diagonal, because  the overlap
of increasingly different
orbitals decay by cancellation of oscillations.
To zero order, we ignore all off-diagonal elements, finding\cite{VOC99}
the small-matrix approximation (SMA)\cite{GKG97}:
\ben
\Omega = \Omega_q + 2\omega_q 
\langle q | f\Hxc (\omega_q) | q \rangle.
\label{OmegaSMA}
\een
The original single-pole approximation (SPA) \cite{PGG96} can be viewed 
as a special case of SMA when the shift from the KS value is small:
\ben
\omega=\sqrt{\Omega_q + 2\omega_q 
\langle q | f\Hxc | q \rangle
}=\omega_q+
\langle q | f\Hxc (\omega_q) | q \rangle
+\ldots
\label{omegaSPA}
\een
In the special case of including just the exchange kernel,
this result is identical to
first-order GL perturbation theory \cite{GS99}.

To go beyond SPA, we use
a continued-fraction method \cite{S82}
for inverting a matrix with a dominant diagonal.  Truncating the
expansion
at second order in the off-diagonal matrix
elements:
\ben
\Omega=\Omega_q\SMA + \sum_{q'\neq q} \frac{4\omega_q\omega_{q'}
|\langle q | f\Hxc (\omega_q\SMA) | q' \rangle|^2}
{\Omega_q\SMA-\Omega_{q'}\SMA}.
\label{CFexp}
\een
This is a key result of this paper, leading to many conclusions.
First, it yields the exact GL perturbative expression to {\em second}-order
in $\lambda$.  Expanding 
$f\Hxc=\lambda f\Hx + \lambda^2 f\c^{[2]}
+\ldots$, we find
\ben
\omega=\omega_q+\lambda \langle q | f\Hx (\omega_q)| q \rangle
+ \lambda^2 \delta \omega^{(2)}_q,
\label{omegaGL}
\een
where the second-order shift consists of four terms:
\bea
\delta \omega^{(2)}_q&=&\langle q | f\c^{[2]}(\omega_q) | q \rangle
+ 2 \sum_{q'\neq q} 
\frac{\omega_{q'} | \langle q | f\Hx (\omega_q) | q' \rangle |^2}
{\Omega_q-\Omega_{q'}}
\nonumber\\
&+&\langle q |  f\Hx(\omega_q)| q \rangle \langle q | \frac{\partial f\Hx}{\partial \omega}
(\omega_q)| q \rangle
- \frac{| \langle q | f\Hx(\omega_q) | q \rangle |^2}{2\omega_q}. \nonumber \\
\label{dw2}
\eea
For the ground-state energy, the
second-order correction has been identified as playing a key
role in constructing accurate functionals,
especially in cases of strong static correlation \cite{SPK00}.
It is likely to play a similar role for excitations, and can
be easily extracted from Eq. (\ref{omegaGL}).  

Second, we may now deduce precisely when SPA (or SMA) is
valid.
Defining the shift from the KS value as
\ben
\Delta \Omega_q=\Omega-\Omega_q,
\label{shift}
\een
we rewrite Eq. (\ref{CFexp}) in the following suggestive form:
\ben
\Delta \Omega_q=\Delta\Omega_q\SMA\left\{1+
\sum_{q'\neq q} \frac{\Delta\Omega_{q'}\SMA}{\Omega_q\SMA-\Omega_{q'}\SMA}
\frac{|M_{qq'}|^2}{M_{qq}M_{q'q'}}\right\},
\label{DOq}
\een
where $M_{qq'}=\langle q|f\Hxc|q'\rangle$.
A simple estimate 
of the size of this correction can be given
by assuming $M_{qq'} \sim {\sqrt{M_{qq} M_{q'q'}}}$.
This would be exactly true if $f\Hxc(\br,\br',\omega)=f(\br,\omega) f(\br',\omega)$, and
is accurate to within 0.5\% for exchange in the He atom.
Then the SMA is valid when
\ben
\sum_{q'\neq q} \frac{\Delta\Omega_{q'}\SMA}
{\Omega_q\SMA-\Omega_{q'}\SMA} \ll 1,
\label{est}
\een
i.e., the SMA shift need only be small on the scale of the
separation between transition frequencies.  Thus, even when
the corrections to KS transition frequencies are {\em large},
SMA can remain valid if the poles are well-separated.

\begin{table}
\caption{\label{t:table1}
Exact results for the He and Be atoms, using numerically exact
ground-state Kohn-Sham potentials.}
\begin{ruledtabular}
\begin{tabular}{c|c|ccc|cc}
Atom&trans.&\multicolumn{3}{c}{frequency (eV)}&
\multicolumn{2}{c} {osc. str.}\\
&&KS&SMA\footnote{Hybrid SPA results from Ref. \cite{BPG00},
converted to SMA.}
&exact&KS&
exact\footnote{He numbers are from 
Ref. \cite{KH84} and
Be numbers from Ref. \cite{Cb98}.}\\
\hline
He&1s$\to$2p&21.15&21.23&21.22&0.3243& 0.2762\\
&1s$\to$3p&23.06&23.10&23.09&8.47(-2)&7.34(-2)\\
&1s$\to$4p&23.73&23.75&23.75&3.41(-2)&2.99(-2)\\
&1s$\to$5p&24.04&24.05&24.05&1.71(-2)&1.50(-2)\\
&1s$\to$6p&24.21&24.22&24.22&9.8(-3)&8.6(-3)\\
\hline
Be&2s$\to$2p&3.61&4.95&5.28&2.5422&1.3750\\
&2s$\to$3p&7.33&7.39&7.46&3.79(-2)&9.01(-3)\\
&2s$\to$4p&8.29&8.31&8.33&2.06(-2)&2.3(-4)\\
&2s$\to$5p&8.69&8.70&8.69&1.08(-2)&8.1(-4)\\
&2s$\to$6p&8.90&8.90&8.90&6.3(-3)&7.5(-4)\\
\end{tabular}
\end{ruledtabular}
\end{table}
In the special case when the SMA correction is
small compared to a KS transition itself, this result
simplifies to
\ben
\sum_{q'\neq q}\frac{
\langle q | f\Hxc | q' \rangle}
{\omega_q- \omega_{q'}} \ll 1.
\label{estw}
\een
For example, the allowed transitions
from the ground-state of the He atom are listed in Table \ref{t:table1}.
Comparing KS transition frequencies with physical ones, we find
them good to within less then 0.1 eV.   This implies all matrix elements
$\langle q| f\Hxc | q' \rangle$ are small, and SPA is valid.
Calculations within ALDA for this case \cite{00CGGG}
show no difference between full solution of the response equations
and the SPA result.  The column marked SMA lists SMA results with our
best estimate of $f\Hxc$ for this case, a hybrid of exact exchange
with ALDA antiparallel correlation \cite{BPG00}.
For the transition to 2p, the exact exchange result is
21.37 eV \cite{PGB00}, showing that there must be substantial
cancellation by correlation effects.  
For this system and others like it, GL perturbation theory
converges slowly, while our expansion converges rapidly.

Our other prototype is the excitations from the ground
state of the Be atom.  For the 2s $\to$ 2p transition,
the expectation value of $f\Hxc$ is relatively large.
We expect SMA
to work quite well for that transition, but less well for
others, since the one strong transition contributes to the
correction in Eq. (\ref{DOq}), especially having a small denominator.
This is born out by the frequency results in Table \ref{t:table1}.
Within ALDA \cite{00CGGG}, the SMA transition is at 5.27 eV, but the full
calculation is 5.08 eV.

In the SMA, in which off-diagonal elements are
neglected, the eigenvalues in Eq. (\ref{mat}) remain unit vectors, and 
the oscillator strengths retain their KS values.
When we include the change due to the off-diagonal
elements to leading order, we find:
\ben
f = \frac{2}{3} \left( \omega_q \bmu_q^2 +\sum_{q'\neq q}
\frac{4
\langle q | f\Hxc | q' \rangle
\omega_q \omega_{q'} \bmu_q \bmu_{q'}}
{\Omega_q\SMA-\Omega_{q'}\SMA}\right).
\label{fexp}
\een
Here $\bmu_{q}$ denotes the KS dipole matrix element.
Our first conclusion from this important result is
that it contains the exact GL expression for oscillator
strengths to first order in $\lambda$, by ignoring the
correlation kernel.
The sum is
rapidly converging, as the matrix element decays
rapidly with principal quantum number.  Using $f\x$ for
the He atom, the corrections of Eq. (\ref{fexp}), summed
over only bound-bound transitions, reduce the oscillator
strength by 11\%, whereas the exact answer is 15\% lower than
the KS value.  The remaining reduction is due to either correlation
effects or transitions to the continuum.

If we estimate off-diagonal elements with geometric
means of diagonal elements, we find
\ben
f\sim f_q \left( 1 + 2 \sum_{q'\neq q}
\frac{{\sqrt{\Delta\Omega_q\SMA\Delta\Omega_{q'}\SMA}}}
{\Omega_q\SMA-\Omega_{q'}\SMA}
{\sqrt{\frac{f_{q'}}{f_q}}}\right).
\label{fsimp}
\een
The effect of off-diagonal matrix
elements is to mix various KS oscillator strengths.
For the dominant transition ,
if Eq.(\ref{est}) is satisfied for
excitation energies, it is also satisfied for oscillator
strengths. 
The correction to an oscillator strength
is {\em first}-order in the off-diagonal matrix element,
as opposed to the second-order correction to the SMA
transition-frequency shift.  Thus, fractional corrections to KS oscillator
strengths will generally be larger than those to SMA shifts.
So even with large SMA shifts,
the associated oscillator strengths can
be good.  

The oscillator strengths of the He and Be atoms
confirm our previous conclusions.  For the well-separated
transitions (He atom), the KS oscillator strengths
are close to the true oscillator strengths, but not
as close as the transition frequencies.  The deviations
estimated in Eq. (\ref{fsimp}) are consistent with
those of the transition frequencies.  
Similarly, in the case of the Be atom, we see that
the 2s$\to$2p KS oscillator strength is good to within a factor of
2, because the corrections due to other transitions
are quite mild.  On the other hand, the higher transitions
have KS oscillator strengths that are an order of magnitude
different from the true ones, because of the huge corrections
due to the first transition.

Ours is not an accurate
numerical solution of the TDDFT response
equations, but is rather a method for understanding 
results.  For example\cite{00CGGG},
the He oscillator strengths within ALDA are good to within
5\%.  The present work shows that this reflects
the accuracy of $\langle q | f\Hxc\ALDA | q'\rangle$
for these transitions.
Our approximation scheme handles well-separated
poles.  As the system-size grows, e.g., even for small clusters,
excitations become more dense on the $\omega$-axis, and kernel
corrections can cause levels to reorder, etc.  We believe that
these methods can still provide estimates of expected trends,
but this remains to be tested.  It is unclear what happens in
the thermodynamic limit, without better approximations to the 
XC kernel\cite{01BKBL}.
For finite systems, the Tamm-Dancoff
approximation neglects all matrix elements between
one forward and one backward transition (reducing the
size of matrices by a factor of 2), and has been
found to often yield excellent results compared to a full
solution\cite{HH99}.  However, this approximation violates the
Thomas-Reiche-Kuhn sum rule, whereas our expansion
satisfies it order-by-order.  

We have focussed exclusively on the
allowed transitions, because we need exact results on both
transition frequencies and oscillator strengths for our
analysis.  To analyze the limitations of SMA for singlet-triplet
splittings\cite{VOC99} 
would require knowledge of the exact oscillator strengths of
spin-flipping transitions.

So far, we have focussed on approximate solutions
of the TDDFT response equations for the exact ground-state
KS potential.  In practice, this potential is approximated.
Local-density and generalized-gradient
approximations have potentials that are too shallow, so that
Rydberg states are not bound.
This can be corrected by some addition
of the correct asymptotic behavior \cite{CS00}, or
by use of an orbital-dependent functional \cite{GKKG99,IHB99,G99},
whose derivative
yields an accurate potential at large distances.
From accurate calculations on the He and Be atoms, we
find the principal effects of using either exchange-only or
LDA-SIC potentials to be a shift in the orbital energies,
numerically identical to the error in  the ionization potential.
But the KS dipole matrix elements are extremely accurate in these
approximate potentials,
so that the dominant error in oscillator strengths comes
from the errors in eigenvalues.
Thus current technology allows accurate calculation of KS
oscillator strengths.

We conclude with an observation that should be useful for
development of approximations to $f\xc$.  In cases
where poles are well-separated and SMA is valid, as can
be determined by comparing oscillator strengths, the difference
between KS transition frequencies and physical ones
yields an exact expectation value of $f\Hxc$, to 
{\em all} orders in coupling-constant $\lambda$.
This would provide an invaluable benchmark for testing 
approximate XC kernels, similar to the widespread use
exact XC potentials have enjoyed in the ground-state case\cite{UG94}.

We thank C. J. Umrigar for providing us with his exact Kohn-Sham
potentials for the He and Be atoms.
H. Appel acknowledges the
support of a fellowship of the Studienstiftung des deutschen Volkes.
This work was supported by the National Science Foundation under grant
no. CHE-9875091, and also partially by the Petroleum Research Fund.

\end{document}